\documentstyle[12pt]{article}

\begin{document}

\title{
Numerical Proof of Self-Similarity in Burgers' Turbulence
}

\author{
Erik Aurell$^{1,2}$, Sergey~N. Gurbatov$^3$ \& Sergey~I. Simdyankin$^{2,3}$\\
}

\date{}

\maketitle

\vfill

\begin{tabular}{ll}
$^1$ & Mathematics Dept., Stockholm University\\
& S--106 91 Stockholm, SWEDEN\\\\
$^2$ & Center for Parallel Computers,\\
& Royal Institute of Technology,\\
& S--100 44 Stockholm, SWEDEN\\\\
$^3$ &Radiophysics Dept., University of Nizhny Novgorod\\
&23, Gagarin Ave., Nizhny Novgorod 603600, RUSSIA \\
&(Permanent address) \\ \\
\end{tabular}

\vfill

\begin{abstract}

We study the statistical properties of solutions to Burgers' equation,
$v_t + vv_x = \nu v_{xx}$, for large times, when the initial velocity and
its potential are stationary Gaussian processes.  The initial power
spectral density at small wave numbers follows a steep power-law $E_0(k)
\sim |k|^n$ where the exponent $n$ is greater than two.  We compare
results of numerical simulations with dimensional predictions, and with
asymptotic analytical theory.  The theory predicts self-similarity of
statistical characteristics of the turbulence, and also leads to a logarithmic
correction to the law of energy decay in comparison with dimensional
analysis.  We confirm numerically the existence of self-similarity for the
power spectral density, and the existence of a logarithmic correction to the
dimensional predictions.

\end{abstract}

\vfill

\centerline{TRITA-PDC Report 1996:2}

\pagebreak

\section{Introduction}

We study the partial differential equation 
\begin{equation}
v_t + vv_x = \nu v_{xx}
\label{BE}
\end{equation}
in the limit of vanishing coefficient $\nu$.
Here $v(x,t)$ can be taken as the velocity of a wave field, a function
of one-dimensional spatial coordinate $x$ and time $t$, and $\nu$ as 
the viscosity of the medium.

This equation, first introduced by J.M.~Burgers \cite{Burgers} as a model
of hydrodynamic turbulence, arises in many situations in physics.  In the
theory of wave propagation Burgers' equation (\ref{BE}) describes the
evolution of nonlinear waves in non-dispersive dissipative media
\cite{Whitham}.  An important example of such waves is nonlinear
acoustic plane waves \cite{RudenkoSoluyan}.  When a random force is added
to the right-hand side it describes surface growth phenomena
\cite{Kardar,Schwartz}.  Burgers' equation is an effective equation on
large scales, when parity-invariance holds, and arises as an
approximation in description of large-scale structures in the Landau-Ginsburg
equation \cite{Kuramoto}.  One further non-trivial application is to model the
formation and evolution of large-scale structures in the Early Universe,
within the so-called adhesion
approximation \cite{Gurbatov,GurbatovSaichev,Shandarin}.

From the equation (\ref{BE}) the following equation for the potential of
the velocity, 
\begin{equation}
v = - \psi_x,
\label{POT}
\end{equation}
is derived:
\begin{equation}
\psi_t = \frac{1}{2} (\psi_x)^2 + \nu\psi_{xx}.
\label{PBE}
\end{equation}
By using the Hopf-Cole transformation,
$\psi = 2 \nu \ln \theta$,
equation (\ref{POT}) is transformed to the
linear diffusion 
equation \cite{Hopf,Cole}:
\begin{equation}
\theta_t = \nu \theta_{xx}.
\label{HE}
\end{equation}  
The solution of (\ref{HE}) is found by standard convolution with 
the heat kernel, which in the limit $\nu \rightarrow 0$ may be 
evaluated by the method of steepest descent.  

The solution of Burgers' equation (\ref{BE}) in the limit 
$\nu \rightarrow 0$ then has the form: 
\begin{equation}
v(x,t) = \frac{x-a(x,t)}{t}
\label{VEL}
\end{equation} 
where $a(x,t)$ is the coordinate where the maximum of the function 
\begin{equation}
G(x,a,t) = \psi_0(a) - \frac{(x-a)^2}{2t}.
\label{Gf}
\end{equation} 
is obtained.
The velocity potential is found as
\begin{equation}
\psi(x,t) = \max_a \left[ G(x,a,t) \right].
\label{MAX}
\end{equation}
In the continuum
there are two equivalent 
ways of computing $v(x,t)$. Either, the velocity may be 
found by taking the derivative of the function $\psi(x,t)$ from 
(\ref{MAX}), using that the velocity is the gradient of the
velocity potential (\ref{POT}), 
or equation (\ref{VEL}) may be used directly.
In a discrete realization the two ways are still equivalent,
provided we make the correct interpolation between the
grid points. For long times the difference is not important,
and will be ignored in the following.

We use the following definition of the spectral density of energy: 
\begin{equation}
E(k,t) = \frac{1}{2\pi} 
\int_{-\infty}^{\infty}
<v(x+s,t)v(x,t)> \mbox{\Large e}^{i k s} \, ds,
\label{PS}
\end{equation} 
and we are interested in the case where the initial data is a 
Gaussian process, with continuous spectral support around
a dominant wave-number $k_0$.
We want to know how the spectrum changes with time,
for sufficiently long times.
Our approach applies when the initial energy spectrum grows
faster than $k^2$ when $k$ is much less than $k_0$,
and decays quickly when $k$ is larger than $k_0$.
More to the point, we consider the case where the initial
spectrum is a power-law at small $k$
\begin{equation}
E_0(k) \sim |k|^n \qquad |k| << k_0
\label{IPS}
\end{equation}
where the exponent $n$ is a real number greater than two.
At large $k$ we have a smooth cut-off.
The construction of the initial conditions including
normalization factors is described below in section~\ref{s:numerical}.

\section{Summary of dimensional predictions}
\label{s:summary}
We assume a dominant initial wave band around $k_0$, which
implies a correlation length of the initial data $l_0\sim 1/k_0$.

From the characteristic length and the fluctuations
of initial velocity we can form a characteristic time $t_{nl}$, the 
nonlinear time or the characteristic time of shock formation: 
\begin{equation}
t_{nl} =
\frac{ l_0 }{\sqrt{ <v_0^2>}}.
\label{TNL}
\end{equation}

As an illustration we show in fig.~1 one typical realization
of the initial conditions, and how it develops in time.
The characteristic scale $l_0$ can be estimated as the average
distance between local maxima (or minima) of the smooth initial
data (fig.~1a). At the nonlinear time the initial data has changed
to a saw-tooth wave with approximatively the same characteristic
scale (fig.~1b).
At larger times the velocity gradient between the shocks decreases,
and the characteristic scale increases (fig.~1c).
This behaviour is intimately related to the change of the spectral
shape with time, to which we now turn.

We will consider in the following only the asymptotic 
solutions of (\ref{BE}) at times much greater than the
nonlinear time,
\begin{equation}
t \gg t_{nl}.
\label{TSS}
\end{equation} 
The characteristic length scale at this time is
derived by balancing in order of magnitude the two terms
in the Hopf-Cole condition (\ref{Gf},\ref{MAX}):
\begin{equation}
\sqrt{<\psi^2_0>} \sim \frac{<(x-a(x))^2>}{t} \sim \frac{l^2}{t}.
\label{dimensional_balance}
\end{equation}
The distances over which the maximizing is effectively taken at 
time $t$ are therefore such that
\begin{equation}
l(t) \sim \left(t \sqrt{<\psi_0^2>}\right)^{\frac{1}{2}},
\label{LT}
\end{equation}
or, in more transparent form,
\begin{equation}
l(t) \sim l_0 (t/t_{nl})^{1\over 2}.
\label{LTalt}
\end{equation}
We therefore have the picture that at time $t$ the solutions
are composed of shocks, separated by a typical distance
$l(t)$. The spatial gradient of velocity between shocks
is $1/t$ from (\ref{VEL}). The typical velocity and the
typical amplitude of velocity jumps across shocks 
is therefore $l(t)/t$.
From these considerations, the mean energy in the velocity
field per unit length at time $t$ is:
\begin{equation}
E(t) = <v^2> \sim \frac{l(t)^2}{t^2} \sim <v_0^2> (t/t_{nl})^{-1}.
\label{EET}
\end{equation}

Qualitatively, it is possible to estimate the behavior of the power
spectral density from the following considerations. As is well known,
the development of shocks leads to the
asymptotic of the spectrum at large wave-numbers, $E(k,t) \sim
k^{-2}$, and the parametric pumping of energy to the area at small $k$'s
leads to the universal quadratic law, $E(k,t) \sim k^2$
\cite{Burgers,Gurbatov}.
Therefore the spectrum should consist of two branches,  smoothly
transformating from one asymptotics $k^2$ to another $k^{-2}$.
At the turnover point, $k^*$, the spectrum has a maximum:
\begin{equation}
E(k,t) \sim  
\left\{
\begin{array}{ll}
A(t)k^2,    & \mbox{ if } k \ll k^* \nonumber \\ 
B(t)k^{-2}, &\mbox{ if } k \gg k^*  \nonumber
\end{array}
\right.. 
\label{SSPST}
\end{equation}
The only possible scale for $k^*$ is
\begin{equation}
k^*(t) \sim \frac{1}{l(t)}.
\label{KT} 
\end{equation}
For the prefactors in (\ref{SSPST}), one obtains
by comparing with the total energy (\ref{EET})
and matching at $k^*$: 
\begin{equation}
A(t) \sim 
<v_0^2> l_0^3 \left( \frac{t}{t_{nl}} \right)^{1/2},
\label{AT}
\end{equation}
\begin{equation}
B(t) \sim \frac{<v_0^2>}{l_0} \left( \frac{t}{t_{nl}} \right)^{-3/2}.
\label{BT}
\end{equation}
We see that (\ref{AT}) and (\ref{BT}) could also have been derived
directly by dimensional analysis, since
$A(t)\sim \frac{l^5(t)}{t^2}$ and
$B(t)\sim \frac{l(t)}{t^2}$.

It should be noted that if $\nu$ is small but finite we can form 
a ratio between the dissipation time over spatial scale
$l_0$ and the nonlinear time, which we can
naturally refer to as the initial Reynolds number.
\begin{equation}
Re_0 = {{l_0^2/\nu}\over {t_{nl}}} = {{l_0 <v_0^2>^{1/2}}\over{\nu}}
\label{Re_time_0}
\end{equation}
Similarly, at time $t$ we can form the ratio between
the dissipation time over spatial scale $l(t)$ and the turn-over
time, which is on the order of $l(t)/\sqrt{<v^2(t)>}$,
that is, simply proportional to $t$:
\begin{equation}
Re(t) = {{l^2(t)/\nu}\over {t}} \sim {{l_0 <v_0^2>^{1/2}}\over{\nu}}
\label{Re_time_t}
\end{equation}
Within dimensional considerations the Reynolds number hence
remains constant, because the increase in $l(t)$ cancels
with the decrease in typical velocity.

Another way to introduce the Reynolds number, is to observe that
the width of a shock
is of the order of $\nu/\sqrt{<v^2(t)>}$.
Reynolds number is thus also equal to the ratio between the
inter-shock distance at time $t$, that is $l(t)$,
and the width of the shock.
Therefore, as long as Reynolds number stays the same,
the Hopf-Cole transformation and approximation that the
solutions have 
saw-tooth shape are relatively equally accurate at arbitrarily large
times.

We will see in the next section that by including
logarithmic corrections to the dimensional
estimates, Reynolds number will in fact slowly decrease
in time. There will be a characteristic time over which
Reynolds number decreases to order unity, and then viscosity
can not be taken small. This time will however be exponentially
large in the square of $Re_0$, so if viscosity is initially weak
it can be taken weak for very long time, although not infinitely long.

\section{Beyond the dimensional predictions}
\label{s:beyond}

For Burgers' turbulence, in the cases that interest us here,
it is possible to develop an asymptotic theory going beyond
dimensional analysis, which
is valid at times much greater than $t_{nl}$.
From the mathematical point of view
we use the property that for large times the parabola
in (\ref{Gf}) varies little over a distance of the
order of $l(t)$ at its minimum. When $l(t)$ is much larger than
$l_0$, and the correlations in the initial conditions decay
sufficiently fast beyond $l_0$, then a large 
number of local maxima of the initial
potential $\psi_0(a)$ competes to be the global maximum of the
function $G(x,a,t)$, and these local minima are practically
independent.

We can then appeal to result in probability theory
on the expected maxima of Gaussian processes \cite{Cramer,Tikhonov}.
A more direct, although more cumbersome,
approach to derive the asymptotics also
exists\cite{FF}, which has in addition the advantage that one may thus also
compute order by order corrections to the leading
behaviour\cite{AurellFrisch1995}.

We begin by observing that the mean square velocity
and mean square velocity potential are both finite for the process
we consider.
By comparing them we can thus form a characteristic length
\begin{equation}
l_0 = \sqrt{{<\psi_0^2>}\over{<v_0^2>}},
\label{l0def}
\end{equation}
which will of course also be of the order $1/k_0$, but with
a well-defined finite prefactor depending on the functional
form of the spectrum.

Let us consider the decay of energy with time.
Averaging both sides (\ref{PBE}) over space leads to
\begin{equation}
\partial_t <\psi> = E(t)
\end{equation}
The decay of energy can thus be derived from the expectation value
of the velocity potential at time $t$.
The maximization position $(a)$ in (\ref{MAX}) can at most be a distance
of leading order $\sqrt{t}$ away from the Eulerian coordinate $(x)$.
The expectation value of the velocity
potential at time $t$ should therefore be similar to the 
expected maximum value of the initial velocity potential on
a stretch of the line of length  $\sqrt{t}$.
One would expect this value to grow in time, although not very fast.
In fact, a well-known result on stationary
Gaussian processes\cite{Cramer, Tikhonov, Gurbatov} states that
this maximum has a doubly exponential distribution, such that
the expected maximum on a stretch of the line of length
$L$ is $<\psi_0^2>^{1/2}\sqrt{2\log {{L}\over{l_0 \sqrt{2\pi}}}}$,
where $<\psi_0^2>$ has the same meaning as introduced
above, and $l_0$ as in equation (\ref{l0def}).

Substituting $l_0(t/t_{nl})^{1/2}$ for $L$
we arrive at
\begin{equation}
<\psi(t)>\, \sim\, \sqrt{<\psi_0^2>}\sqrt{\log {{t}\over{2\pi t_{nl}}}}
\label{psi_decay}
\end{equation}
Although we should not a priori expect that (\ref{psi_decay})
is correct up to constants, it
turns out to be the case \cite{Kida,FF,GS81,AurellFrisch1995}. 
By differentiating in $t$ follows Kida's law
for the the deacy of energy at long times:
\begin{equation}
E(t) = \frac{1}{2} <v_0^2> (t/t_{nl})^{-1}
\left(\log (t/2\pi t_{nl})\right)^{-1/2}
\label{E_decay_log_corrections}
\end{equation}
Again, this expression holds in the asymptotic theory up to
additive corrections that become small relative to
(\ref{E_decay_log_corrections}) in the limit.

We can also estimate the behaviour of the
characteristic length.
When time grows, the expected maximum of the term $\psi_0$ in (\ref{MAX}) will
slowly become larger than $(x-a)^2/2t$, if the effective length
of the maximization operation, $(x-a)$, is just taken
to be of order $\sqrt{t}$. The balance equation,
(\ref{dimensional_balance}), 
therefore slowly becomes inappropriate to fix the spatial
scale.
Let us therefore instead 
assume that the maximum is attained at a value of $(x-a)$ around $L$.
The attained value of $\psi(x,t)$ would then typically be
\begin{equation}
<\psi>_{L}\, \sim\, \left(
\sqrt{<\psi_0^2>}\sqrt{2\log {{L}\over{l_0 \sqrt{2\pi} }}}
\right) - {{L^2}\over {2t}}.
\label{MAX_l_of_t}
\end{equation}
If we now  maximize (\ref{MAX_l_of_t}) over  $L$ we find
that the maximum is attained at a value of
\begin{equation}
l(t) \sim l_0 (t/t_{nl})^{1/2}(\log {{t}\over{2\pi t_{nl}}})^{-\frac{1}{4}}
\label{l_decay}
\end{equation}
The characteristic length-scale therefore grows slightly more slowly than
the dimensional estimate.
Let us note that the estimate of the maximum (\ref{psi_decay}) remains
unchanged. Only subleading terms are influenced by the small correction
to the characteristic length.
Let us also note that the dimensional estimate
$E(t)\sim l^2(t)/t^2$ remains correct up to logarithmic corrections,
which indicates that the fluctuations are very small.

In a similar spirit one may also compute 
two-point probability densities \cite{Gurbatov,GS81},
correlation functions and the power spectral density
\cite{Kida,FF,GS81, Gurbatov}. 
In particular, the energy spectrum has the expression 
\begin{equation}
E(k,t) = \frac{l^3(t)}{t^2}\tilde E(k l(t)),
\label{PST}
\end{equation}
where $l(t)$ is the characteristic scale in (\ref{l_decay}),
and the function $\tilde E(\kappa)$
can be computed to the same leading order in the
asymptotics\cite{Gurbatov,GS81,Kida}.

In fig.~2~(a) three power spectra 
(averaged over 1000 realizations of the random process) 
are shown at three different moments of time. 
The initial spectrum was $k^3$ at small $k$.
The preservation of the shape of each curve is evident. It is also seen
that the total energy (the area under the curves) decreases, and so does
the characteristic wave-number $k^*$.
Fig.~2~(b) contains three power 
spectra computed at the same time at the self-similar stage 
(averaged over 1000 realizations). 
The initial spectra were
$k^3$, $k^6$ and $k^{12}$ at small $k$, respectively.

The analytical expression for $\tilde E$ is somewhat
unwieldy.
However, qualitatively, $\tilde E$ is simply described as a smooth
interpolation between two power-law asymptotes that 
hold when $\kappa$ is respectively much smaller or much
larger than unity.
In the first case
$\tilde E(\kappa) \sim 1.08 \kappa^2$, in agreement with the expected leading
$k^2$ behaviour of the spectrum
according to (\ref{SSPST}), while for large $\kappa$
$E(\kappa) \sim 0.359/\kappa^2$, also
in agreement with (\ref{SSPST}) \cite{Gurbatov}.
In Fig.3 $\tilde E(\kappa)$ is shown together with numerical
results obtained by averaging over 1000 realizations of the
initial conditions.
The numerically obtained curve 
(as a result of averaging over 1000 realizations of the initial 
random process) is artificially shifted slightly 
downwards to be distinguishable from the analytical curve.

It is seen that the asymptotic formula describes the numerical
data very well, not only in the limits of relatively
large and small wave numbers, but also at the top, where
the spectrum switches between the two asymptotes.
From fig.~3 is also seen that the
transformation region from the top down towards
$\kappa^{-2}$ is thin, but the transformation region
from the top down to $\kappa^2$ is considerably wider.

In fig.~4 the function $l(t)$ is shown.  
The dashed line corresponds to
the calculation based on the expression (\ref{LT}). The result of the
computations taking into account the logarithmic correction 
(\ref{l_decay}) is shown by the solid line. 
The centered symbols show results from 
numerical simulations of an ensemble of 500 realizations.
The length scale was estimated from the numerical
data in the following way: the averaged spectrum was first
found, and by fitting the two asymptotes 
$Ak^2$ and 
$Bk^{-2}$,
their intersection was taken as the main wave number $k^*$.
The numerical estimate of $l(t)$ was then defined as $1/k^*$.
The fact that the results with the logarithmic  corrections fit the
data better proves numerically the validity of the
asymptotic theory.

In fig.~5  we show the results of a procedure analogous to that depicted in
fig.~4 but for the function $E(t)$. It is seen again
that the solid line obtained by taking into account the logarithmic
correction fits the numerics (centered symbols) 
better. Since the correction is slow, the difference between this result
and the result of the dimensional analysis (dashed line) is
visible but small.

From which have been shown above one might conclude
that the logarithmic correction is not very significant, since
it gives a small addition to the dimensional predictions.
To a certain extent that depends
on what quantity one looks at.
We can for instance compute the prefactors of the power-law
for the spectrum at large and small wave numbers.
If $E(k)\sim A(t) k^2$ up to a maximum wave number $k^*$,
then the total energy at time $t$ is to leading order
$A(t) (k^*)^3$. But we know already the energy and $k^*$
as functions of $t$, so by comparing we compute $A(t)$.
By matching at $k^*$ we then also find $B(t)$, and the
results are:
\begin{equation}
A(t) \sim 
<v_0^2> l_0^3 \left( \frac{t}{t_{nl}} \right)^{1/2}
\left( \log {{t}\over{2\pi t_{nl}}} \right)^{-\frac{5}{4}}
\label{ATl}
\end{equation}
\begin{equation}
B(t) \sim
\frac{<v_0^2>}{l_0} \left( \frac{t}{t_{nl}} \right)^{-3/2}
\left( \log {{t}\over{2\pi t_{nl}}} \right)^{-\frac{1}{4}}
\label{BTl}
\end{equation}
In the case of $A(t)$, the correction to the dimensional predictions
is more significant than those shown previously.
In fig.~6 the centered symbols 
show the values of $A(t)$ obtained from the numerical simulation.
The dashed line is the dimensional prediction computed according to (\ref{AT}),
while the solid line takes into account the logarithmic correction 
(\ref{ATl}).

There is also a real physical interest in the logarithmic
corrections, since they change qualitatively the behaviour
at large times. Let us form the Reynolds number at time $t$,
as in (\ref{Re_time_t}):
\begin{equation}
Re(t) = {{l(t)\sqrt{E(t)}}\over{\nu}} \sim 
Re_0
\left( \log {{t}\over{2\pi t_{nl}}} \right)^{-\frac{1}{2}} 
\label{Re_time_t_log}
\end{equation}
where $Re_0$ is the initial Reynolds number as in (\ref{Re_time_0}).
Although we compute the solutions to Burgers' equation in the
limit where viscocity vanishes,
that is, in the infinite Reynolds
number limit, Reynolds number still decreases with time
\cite{Burgers,Gurbatov,GS81}. 
Sooner or later we will reach a stage where viscosity can not
be taken arbitrarily small, and at even later time we will
eventually reach a the linear stage of decay.
The time to reach the linear regime can be estimated
by taking $Re(t_l) \sim 1$, which leads to   
\begin{equation}
t_l \sim 2\pi  t_{nl} \exp(Re_0^2).
\end{equation}

\section{Different dimensional ans\"atze}
\label{s:different}
We started our more precise analysis
by expressing the correlation
length as $l_0 =\sqrt{{<\psi_0^2>}\over{<v_0^2>}}$.
We might also consider the initial
velocity gradient, and its variance, that we
write $\sigma_u^2 = <(\partial_x v_0)^2>$.
We could then define the correlation
length by $l_0' = \sqrt{{<v_0^2>}\over{\sigma_u^2}}$,
and a new nonlinear time by
$t_{nl}' = 1/\sigma_u$.
This definition of the nonlinear time has some
appeal, since a shock first forms where the velocity
gradient has a local minimum, and does so after 
a time $1/|\partial_x v_0|$.

For initial spectra going as $k^n$ with $n$ greater
than two, $l_0'$
and $t_{nl}'$ are of course simply proportional
to $l_0$ and $t_{nl}$.
In dimensional analysis it does not matter which ones
we use, but, as we have seen, in the asymptotic
analysis the correlation length and the nonlinear
time that appear are those given above and used in
section~\ref{s:beyond}.
Furthermore, a considerable part of the asymptotic
analysis is valid down to $n$ equal to one
\cite{AurellFrisch1995}.
In that limit $l_0$ diverges, while $l_0'$ remains finite.
The  expressions from dimensional analysis
using $l_0'$ and $t_{nl}'$ therefore grow progressively less accurate
as $n$ tends to one from above.

If we lower $n$ further and consider the interval $[-1,1]$,
$l_0$ and $t_{nl}$ do not exist, because
the initial velocity potential is then  not a stationary process,
only a process with stationary increments.
The initial velocity will still be stationary however,
and its correlation length will be approximatively 
$l_0'$, and
the shock formation time will still be about
$t_{nl}'$.
For distances $|a-x|$ much greater than $l_0'$,
$<(\psi_0(a) - \psi_0(x))^2>$ is equal to $ \beta^2
  (|a-x|)^{1-n}$,
where $\beta^2$ is given as
$\alpha^2 {{4\pi}\over{\Gamma(2-n)\cos\frac{\pi n}{2}}}$.
Balancing in (\ref{MAX})  gives
$l(t) \sim (\beta t)^{\frac{2}{3+n}}$.
We see that $\beta^2$~diverges as $n$ tends to one from below.

If we instead proceed by dimensional analysis using $l_0'$, then
$<(\psi_0(a) - \psi_0(x))^2>$
should be similar to $ (l_0')^2<v_0^2> ({{|a-x|}\over{l_0'}})^{1-n}$,
with a numerical factor which depends on how the cut-off
is made at large $k$.
Balancing in~(\ref{MAX}) gives
$l(t) \sim l_0' \left( t/t_{nl}'\right)^{\frac{2}{3+n}}$.
The upshot of this discussion is that
$t_{nl}'$ is a good measure of shock formation
time, and can be used as a nonlinear time in the
dimensional estimates in the whole interval of $n$ greater than $-1$,
except around one, where numerical prefactors diverge, because the
long-term asymptotics changes qualitatively around this point.

\section{Numerical work} 
\label{NumWork}

\subsection{Normalizations of initial spectrum}
\label{s:numerical}

We use the following smooth cut-off of the initial power spectrum:
\begin{equation}
E_0(k) = \alpha_n^2 |k|^n
\mbox{\Large e}^{-\frac{k^2}{2k_0^2}}.
\label{NIS}
\end{equation}
We let $k$ vary in the first Brillouin zone, that is
$k$ goes between $-1/2$ and $1/2$, with $k_0$ of the
order of $1/2$. The initial spatial scale
$l_0$ is therefore of order one.

The variances of the velocity and the velocity potential
can  be computed to be:
\begin{eqnarray}
<v_0^2> & = & \alpha_n^2 
\Gamma\left( \frac{n+1}{2} \right) k_0^{n+1},  \\
<\psi_0^2> & = & \alpha_n^2 
\Gamma\left( \frac{n-1}{2} \right) k_0^{n-1}, 
\end{eqnarray}
where $\Gamma(x)$ is the gamma-function. 

Furthermore, we use the convention that $<\psi_0^2>$
is equal to one, which is of course equivalent to
making a choice of scale of time.
We have then
\begin{eqnarray}
<v_0^2> & = & {{ \Gamma\left( \frac{n+1}{2} \right) k_0^{2}  }\over
{\Gamma\left( \frac{n-1}{2} \right) }}
\end{eqnarray}
and, following the above,
\begin{eqnarray}
l_0  & = & \frac{1}{k_0} \sqrt{
{\Gamma\left( \frac{n-1}{2} \right) } \over
 {{ \Gamma\left( \frac{n+1}{2} \right)   }}}
\end{eqnarray}
and
\begin{eqnarray}
t_{nl}  & = & {
{\Gamma\left( \frac{n-1}{2} \right) } \over
 {{ \Gamma\left( \frac{n+1}{2} \right) k_0^2  }}}.
\end{eqnarray}

We can also compute the variance of the velocity gradient,
which is
\begin{eqnarray}
\sigma_u^2 & = & \alpha_n^2 
\Gamma\left( \frac{n+3}{2} \right) k_0^{n+3}, \nonumber \\
\end{eqnarray}

We can hence form the second estimate of the initial
correlation lengths, $l_0'$, and the ratio between
the two lengths, which is
\begin{equation}
\frac{l_0'}{l_0}  = 
\frac{ \Gamma\left( \frac{n+1}{2} \right) }{
\sqrt{
\Gamma\left( \frac{n-1}{2} \right) 
\Gamma\left( \frac{n+3}{2} \right)
}
}, 
\label{GAM}
\end{equation}
We see indeed that these two lengths only differ by a constant
depending on $n$, as was stated in the previous discussion.

\subsection{Generation of initial conditions}

Fourier components of a Gaussian process are independent Gaussian 
variables. We therefore synthesize the initial 
potential of the velocity by first 
generating random variables $a_k$ distributed according to 
\[
P(a_k) = \frac{1}{N} \exp \left( -\frac{a_k^2}{2\sigma_k^2} \right)
\]
where
\[
\sigma_k^2 = E_{\psi_0}(k) dk.
\]
Here is used the well known relation between the power spectra of the
process and its derivative
\begin{equation}
E_{\psi_0}(k) = k^{-2} E_0(k)
\label{SPV}
\end{equation}
where the form of $E_0(k)$ is chosen with a smooth cut-off at large $k$ 
according to (\ref{NIS}).
By an inverse Fourier transform we find the initial potential in 
real space, and repeating the whole process many times we sample
the desired ensemble of Gaussian initial conditions.

\subsection{Fast Legendre Transforms}

In numerical simulations the initial data are always generated as  
a discrete set of $N$ points. It could be assumed naively that 
the number of operations necessary to compute the maximization 
(\ref{MAX}) for all values of $x$ scales as $O(N^2)$. It may however be shown, 
using (\ref{MAX}) that $a(x)$ is a nondecreasing function of $x$.
The number of operations needed in an ordered search therefore scales 
as $O(N\log_2 N)$ when using the so-called Fast Legendre Transform
procedure \cite{SheAurellFrisch,Noullez}.

When the number of grid points is large they can not all fit 
into the working memory of the CPU at the same time.
Even if the operations count is the same, it is preferable 
to limit paging of data from an external memory storage to a minimum. 
This is achieved in the in-order algorithm of Noullez \cite{Noullez} 
that we use here.

\subsection{Discrete Fourier analysis}

After Legendre transform, we have at our disposal
a function $\psi(x,t)$ sampled at $N$ discrete
points 
$\psi(x_l,t)$, $l=1,2,...,N$. The sample interval
is $\Delta x = 1$.
Computing the discrete Fourier transform
by using FFTs as
\begin{equation}
\hat\psi(j,t) = \sum_l\psi(x_l,t) {\Large e}^{2\pi i j l/N}\quad
j = 0,\ldots, N-1
\end{equation}
we will get  Fourier coefficients at wave numbers
$k_j = \frac{\Delta x j}{N}$, where the last half
indicates negative frequencies.
Since the signal is real the Fourier amplitudes 
are related by  
$\hat \psi(k_j,t) = \hat \psi^*(-k_j,t)$. 

Knowing the spectral density of the potential $\hat \psi(k_j,t)$, it 
is easy to find the spectral density of the velocity,
since
\[
\hat v(k_j,t) = ik_j \hat \psi(k_j,t).
\]

Different definition can then be made of the
numerically obtained power spectrum, which only
differ by prefactors of $N$. 
We use here the normalizations given
in a standard reference\cite{NumRec}:
\begin{eqnarray}
E(0,t)   & = & \frac{1}{N^2}|\hat v(0,t)|^2, \nonumber \\
E(k_j,t) & = & \frac{2}{N^2}|\hat v(k_j,t)|^2, \; 
j=1,2,...,\left( \frac{N}{2}-1 \right), \\
E(k_c,t) & = & \frac{1}{N^2}|\hat v(k_c,t)|^2. \nonumber
\label{PSE}
\end{eqnarray}

Since the initial potential $\psi_0(x_l)$ is obtained by an 
inverse Fourier transform it is a periodic function with period $N$.
But the use of the fast Legendre transform of a periodic function 
introduces undesirable edge effects which do not allow to achieve 
periodicity in the resulted potential $\psi(x_l,t)$. This causes 
distortion of the spectrum connected with the finite, but not periodic,
function in real space. We eliminate this effect by constructing 
the periodic potential at time $t$ for the cost of using the fast 
Legendre transform over additional $N$ points of the initial potential. 
The `new' periodic initial potential in obtained by simple addition a 
fragment of $\psi_0(x_l)$ with $l$ in $[N/2+1,N]$ to the left-hand side of 
the same realization. A fragment with $l$ in $[1,N/2]$ is then added 
to the right-hand part. Doing this we again obtain a periodic 
function $\psi_0'(x_l)$ but now with `double' periodicity 
\begin{eqnarray}
\psi_0'(x_l)  = & \psi_0(x_{l + N/2}), & 1 \le l \le N/2, \nonumber \\
\psi_0'(x_l)  = & \psi_0(x_{l - N/2}), & N/2 + 1 \le l \le N+N/2, \nonumber \\
\psi_0'(x_l)  = & \psi_0(x_{l - N - N/2}), & 
N + N/2 + 1 \le l \le 2N. \nonumber
\end{eqnarray}  
After performing the fast Legendre transform over $2N$ points we get a function
$\psi'(x_l,t)$. This function loses the `double' periodicity, 
but the middle $N$ points of this realization
\[
\psi(x_l,t) = \psi'(x_{l + N/2}), 1 \le l \le N
\]
form a periodic function $\psi(x_l,t)$ with period $N$ which we use 
further for the Fourier analysis.  

We find that to check the described theory not very large simulations 
are needed. In Fig.1 -- 6, shown above, we have used $N$ equal to 
$2^{15}$, i.e. $32768$. 
No qualitative difference appears with the use of larger $N$ 
or greater number of realizations. 

\section{Acknowledgments}

This work was supported by the Swedish Natural Science
Research Council through grant S-FO-1778-302 (E.A.),
by the KTH CITEC Project No. 930-131-95 (S.N.G.), and
by the Swedish Institute (S.I.S.). 
E.Aurell and S.I.Simdyankin thank the 
Center for Parallel Computers for hospitality.

\pagebreak

\noindent {\bf Figure captions:}

\begin{itemize}

\item[Figure 1:] 
A realization of the velocity field at time $t=0$ (a), 
evolved process at dimensionless time $t/t_{nl} \sim 10^5$ (b), 
and the same realization at a larger time $t/t_{nl} = 10^6$ (c).  

\item[Figure 2:] 
(a) Power spectral densities computed at different times: 
(1) $t/t_{nl} \sim 5 \cdot 10^8$, (2)  $t/t_{nl} \sim 3 \cdot 10^9$, 
(3) $t/t_{nl} \sim 2 \cdot 10^{10}$. Log-log plot. \\
(b) Three power spectral densities computed at the same time.
The exponent $n$ of the initial power spectrum for the curve above 
$n=3$, and for the other two, which are hardly distinguishable, 
$n=6$ and $n=12$. Log-log plots.	

\item[Figure 3:] 
Power spectral densities resulted from evaluation of an
analytical integral expression (upper curve) and from the direct
numerical simulation (lower curve shifted down to be distinguishable from
the upper one). Log-log plot. 

\item[Figure 4:] 
Characteristic space-scale as a function of time. 
The prediction $l(t)\sim \sqrt{t}$ (dashed line) 
is by dimensional analysis, 
while the more precise approximation is
$(\log t)^{-1/4}\sqrt{t}$ (solid line). Log-log plot.

\item[Figure 5:] 
Total energy as a function of time $E(t)$. \\
Centered symbols - the numerically obtained values of $E(t)$.   
The prediction $E(t)\sim1/t$ (dashed line) is by dimensional analysis, 
while the more precise approximation is
$1/t\sqrt{\log t}$ (solid line). Log-log plots.

\item[Figure 6:] 
The asymptotic prefactor $A(t)$ at the IR part of the 
power spectral density as a function of time. 
Centered symbols - values of $A(t)$ resulted from 
the numerical simulation. 
Dashed line - the dimensional prediction, and 
solid line - analytical result with taking into account 
the logarithmic correction.
\end{itemize}  

\end{document}